# Quartile Clustering: A quartile based technique for Generating Meaningful Clusters

S. Goswami, Dr. A. Chakrabarti

**Abstract**—Clustering is one of the main tasks in exploratory data analysis and descriptive statistics where the main objective is partitioning observations in groups. Clustering has a broad range of application in varied domains like climate, business, information retrieval, biology, psychology, to name a few. A variety of methods and algorithms have been developed for clustering tasks in the last few decades. We observe that most of these algorithms define a cluster in terms of value of the attributes, density, distance etc. However these definitions fail to attach a clear meaning/semantics to the generated clusters. We argue that clusters having understandable and distinct semantics defined in terms of quartiles/halves are more appealing to business analysts than the clusters defined by data boundaries or prototypes. On the same premise, we propose our new algorithm named as quartile clustering technique. Through a series of experiments we establish efficacy of this algorithm. We demonstrate that the quartile clustering technique adds clear meaning to each of the clusters compared to K-means. We use DB Index to measure goodness of the clusters and show our method is comparable to the EM (Expectation Maximization), PAM (Partition around Medoid) and K Means. We have explored its capability in detecting outlier and the benefit of added semantics. We discuss some of the limitations in its present form and also provide a rough direction in addressing the issue of merging the generated clusters.

———————————— ◆ ————————————

## 1 INTRODUCTION

Clustering is one of the very basic type of data analysis and exploration tasks. It partitions data into groups based on similarity or dissimilarity of the members. The groups should be such that similarity between members within same cluster should be high and between clusters it should be low. Unlike classification it does not need a priori information; hence it is known as an unsupervised method. The most important use of clustering is in gaining understanding of the data and data summarization. As a result clustering has its application across board application domains like climate, business, information retrieval, biology, psychology to name a few. Often it is used as a pre cursor to other data analysis tasks as well. As example if a business analyst is trying to find the churn rate for a tele-communication domain, he might be interested to find churn among the young and heavy usage group first. Clustering is also one of the principal methods in outlier detection, which has found extensive use in data cleaning, telecommunication and financial fraud detection [1]. The popularity of this particular task is also endorsed by the fact that clustering is present in most of the leading data mining tools like [2] SAS, SPSS, Weka, SSAS, R. It also observes, clustering is the third most used algorithm in the field of data mining after decision tree and regression. 69% Consultants have voted for decision tree, 68% for regression, 60% for clus-tering. The next task is Time series with 32% votes.

Data mining has been commercialized enough and it has proved to be an invaluable weapon in business analyst's arsenal. Several data mining packages are being used by business and technical communities across organization to unearth meaningful information from structured and unstructured data. One important observation is that not many algorithms have found a place in these packages which are being widely used. Few plausible explanations for the same may be 1) implementation challenges associated with few of the algorithms, 2) it is not generic enough to be applied across diverse application domain 3) the clusters generated are not intuitive to business analysts.

One of the challenges in clustering is associating a cluster with a clear meaning or concept. [3], [4] observes what each cluster stands for, becomes very hard to define or become very subjective [5]. This certainly makes use of the data and devising any particular business/promotional strategy for a cluster way too difficult especially in the commercial domain. A definition of data boundary or a prototype value, which is arrived at by an objective like minimizing sum of square errors, may not make practical sense to the business community. Let's look at an arbitrary example, say the result of a clustering algorithm represent the cluster by a center (25732, 27.63) where the values represent salary and age of a customer. The meaning of the cluster is not very clear to the business analysts. Contrary to that, QueryClustering will generate a cluster which will consist of customers whose salary is in the top 25% and age in the bottom 25%.

Our main contributions in this paper are as follows
1. Proposing an algorithm which attaches a clear

• *S. Goswami is a research scholar with A.K.Choudhury School of I.T., University of Calcutta, Kolkata, India.*
• *Dr. A. Chakrabarti is with A.K.Choudhury School of I.T., University of Calcutta, Kolkata, India.*

meaning/semantics to each cluster, which has its basis of median and quartiles of data.
2. Establish through empirical case studies
3. In terms of "goodness of Cluster" this algorithm is comparable with other algorithm.
4. It brings a better meaning/identification of each cluster
5. How grouping defined in terms of quartiles help in a task like outlier detection

The organization of the paper is as listed below. Section II details the related and recent works in this area. Section III introduces clustering, outlier detection, quartiles very briefly. We also introduce basic working of few popular algorithm and couple of indices. These indices help in comparing the algorithms in terms of goodness of the clusters. In Section IV, we introduce our algorithm which is based on median and quartiles. Section V comprises of three experimental case studies. In section VI we summarize our findings and provide a rough direction in addressing the most important of these (merging cluster). Section VII lists the conclusion, challenges and road ahead.

## 2 RELATED WORKS

Clustering is one of the long studied problems because of its wide applicability. Listing all of the related research works exhaustively is difficult. The algorithms can be classified as partition-based or hierarchical depending on the natures of clusters generated. Based on the methods they can be classified as Partitioning relocation methods, density based, grid based, model based to name a few. Some algorithms are tuned to handle high dimension and high volume data, while some algorithms are designed to handle categorical attributes as well. Few landmark algorithms can be referred at [6], [7], [8], [9], [10], [11], [12], [13]. Our method is closest to grid based methods, in terms of approach. [5] Lists few emerging trends in data clustering as ensemble, semi-supervised clustering, large scale clustering and multi-way clustering.

While we explore works done on semantics /meaning of a cluster we come across research work [14] where clustering has been used to generate a hierarchical metadata from annotations in context of semantic web. In [15] the authors elaborate on use of semantics in case of text clustering. Our focus is more to increase the meaning of the clusters.

There are research works in various fields which talk about quartiles in conjunction with clustering. There is an interesting use of quartiles in [16] which employ quartiles to detect money laundering through inflated export price. Use of quartiles is quite common on medical science. [27], [28] are couple of references on the same.

## 3 PRELIMINARIES

In this section we start with defining basics of clustering. We very briefly touch upon outliers and quartile. Next,

we introduce few indices to evaluate clustering algorithms. We compare results with few clustering algorithms. As a pretext we give a very brief outline of those algorithms

### 3.1 Clustering

In [18] the author defines clustering or cluster analysis in the following way:-

Cluster analysis is a set of methodologies for automatic classification of samples into a number of groups using a measure of association, so that the samples in one group are similar and samples belonging to different groups are not similar. The input for a system of cluster analysis is a set of samples and a measure of similarity (or dissimilarity) between two samples.

To compare the similarity between two observations, distance measure is employed. To express the relationship formally, we reference [13]. If there are two clusters C1 and C2, if observations (p11, p12, p13, ....., p1n) belong to C1 and (p21, p22, p23,......, p2m) belongs to C2 then generically Dist(p1i, p1j)<Dist(p1i, p2k), Where $i{\neq}j$ and 1<=i<=n, 1<=j<=m and 1<=k<=m. Dist is a function which takes two observations and uses any standard distance measure technique and gives the distance between them as output. Some of the popular algorithms are K-Means, Nearest Neighbor, PAM etc. For larger data sets, sampling based approaches like CLARA or CLARANS is used. For a very apt review on the clustering techniques, we can refer to [3], [5], [19], [20]. There had been numerous algorithms developed in clustering. Broadly they can be classified as partition based and hierarchical. The hierarchical algorithm can be further classified as agglomerative and divisive. There are some finer types like density based, graph based or grid based. Some algorithms can handle categorical attributes where as some of them have ability to handle high dimensional data. For some of the algorithms the membership to clusters is crisp whereas for some of them it is fuzzy, i.e. same member can belong to multiple clusters.

We define two terms: Sum of squared error and Davies–Bouldin index to compare our algorithm vis-à-vis other algorithms. We have used the parameter as half, so the number of clusters is four. We have used the same cluster number as input for the other algorithms also. The tools that we have used are SQL Server Analysis Service (SSAS) and 'R' respectively.

### 3.2 Sum of Squared Error (SSE)

The squared error is the sum of Euclidean distance for each observation from the center for a particular cluster. Sum of squared error is the sum of the squared error across clusters as the name implies.

$$SSE = \sum_{i=1}^{n} \sum_{j=1}^{m} ( x_{kj} - C_i )$$

This is the objective function for K-means, which achieves a local optimization rather than a global one. In the above equation k is the no. of clusters and m is the no. of observations in each cluster. $C_i$ represents the center of the Kth cluster.

Also QuartileClustering does not create centers naturally, so we take simple arithmetic mean of the members to arrive at a center.

### 3.3 Davies–Bouldin index

This index is given by

$$DB = \frac{1}{n} \sum_{k=1}^{n} \max_{i \neq j} \left( \sigma_i + \sigma_j \right) / d(Ci + Cj)$$

where n is the number of clusters, cx is the centroid of cluster x, σx is the average distance of all elements in cluster x to centroid cx, and d(ci,cj) is the distance between centroids ci and cj. Since algorithms that produce clusters with low intra-cluster distances (high intra-cluster similarity) and high inter-cluster distances (low inter-cluster similarity) will have a low Davies–Bouldin index, the clustering algorithm that produces a collection of clusters with the smallest Davies–Bouldin index is considered the best algorithm based on this criterion.

We have compared performance of our algorithm against few clustering algorithms. We enclose a brief description of them below.

### 3.4 K-means

One of the most used algorithm because of its simplicity. This is an iterative algorithm where K is the number of clusters. It starts with a random selection of K points and then repeatedly tries to minimize the SSE. The stopping or convergence criteria we can also use as a percentage inter cluster movements that are taking place. Weakness of this algorithm is that it might miss the global optima. It is good for finding convex shaped clusters and do not handle outliers well. For a detailed account we can look at [13].

### 3.5 PAM

Partitioning around medoids also works in an iterative mode. While K-means cluster is represented by a centroid, which is the arithmetic mean and not an actual point, in case of PAM a cluster is represented by a medoid, which is a centrally located point in each cluster. So in each step here also SSE is minimized. It is more robust to outliers than K-means. Its weakness lies in its time complexity. For a detailed account we can look at [23].

### 3.6 CLARA

It is based on the same premise as PAM, It improves the time complexity by working on samples. For a detailed account we can look at [23].

### 3.7 EM

This is a parametric method where each cluster corresponds to a Gaussian distribution. The algorithm iterates through an expectation (E) and Maximization (M) step. In E step expectation of the log-likelihood is evaluated using current parameter. In M step it computes parameters by maximizing the log likelihood as calculated in earlier step.

### 3.8 Quartiles

Quartiles are a major tool in descriptive analysis, which divides the range of data in four parts each having 25% of the data. Q1 is the 25% point, Q2 which is also same as Median, is the 50% point and Q3 is the 75% point. One of the very common uses of quartile can be found in box-plot analysis as depicted in figure 1.

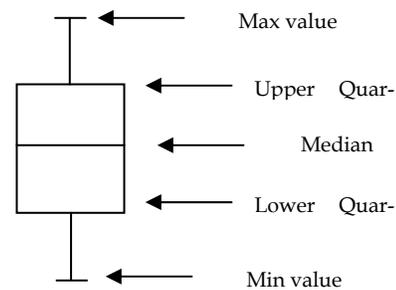

**Figure 1: Box plot**

The spread and central value of any variable can be represented using box-plot.

### 3.9 Outliers

In [18] the authors define an outlying observation, or outlier, is one, that appears to deviate considerably from other members of the sample in which it occurs. Another definition as observed in [21] is, it is an observation that deviates so much from other observations so as to arouse suspicion that it was generated by a different mechanism. For a detailed overview on the outlier we can refer [1], [22]. Outlier is also referred as anomaly, novelty, exception, surprise etc. There are different ways of detecting outlier namely classification based, Nearest Neighbor based, clustering based and statistics based etc. Nearest Neighbor can be further classified in distance based and density based techniques. The statistics based techniques are either parametric (assumes a data distribution model) or non-parametric. They are applied in various domains like fraud detection, intrusion detection, medical data, sports, novel topic detection etc.

## 4 OUR PROPOSAL

We propose a clustering algorithm based on quartiles and median. Let us say, we have a dataset D = {D1, D2, ….. Dn} where each data elements again consist of k attributes i.e. of the form (X1, X2,…XK). As the first step, we map each attribute of each observation to a literal Q1, Q2, Q3, Q4 or H1, H2 based on the input we are providing to the algorithm. For a simplistic example let's consider a dataset { (1,3),(2,4),(3,2),(4,7),(6,9)} which has five observations and each observation is having two attributes. We represent X1 =

{1,2,3,4,6} and X2 ={3,4,2,7,9}. M1 = 3 and M2= 4, where M1, M2 are the median values for X1 and X2 respectively. So for any value which is less than or equal to the median value will be represented as H1 and greater than median as H2 respectively. So if we look back, the first observation (1, 3) will be now H1H1. So basically, each observation gets mapped to one of these groups. There can be a total of 22 groups as we are considering halves, so if we are considering n attributes for each observation and quartiles, the number of groups that are possible are 4n. Our proposition is that all these individual groups correspond to a cluster. Rather than a centroid or a medoid, the cluster is represented by this literal string. It is evident that the number of clusters can be very high for high dimensionality. We discuss a merging technique for the same in the following section. The philosophy is to have a uniform and understandable semantics attached with each cluster, rather than a central value and a distance metrics. We name the algorithm as QuartileClustering as the objective of this method is to find clusters from a dataset based on quartile value of the attributes.

## 4.1 Algorithm QuartileClustering

Input: Dataset, MedQuartIndicator, MergeIndicator
Output: Clustered dataset

Begin
Step 1: Calculate Median/ Quartiles for each attribute based on the value of MedQuartIndicator
Step 2: For each observation
For each attribute
Step 2a: Compare value with either Median or Quartile based on the value of MedQuartIndicator
Step 2b: As per the comparison result convert the numeric value to a literal value
Step 3: Group observations by the literal value
Step 4: if the value of MergeIndicator is "False" Stop
Else merge()*
End

Merge() is a method which would be developed as a future course of work. This parameter has been included in the method signature to support future extension.

This can be done in two passes. In the first pass, we can calculate the medians /quartiles, second pass we convert them to literal values and group them. As a result the algorithm will have only linear time complexity as opposed to a density or a neighborhood based algorithm.

**Example:** We take a dataset having nine observations. Each observation has two features. Let the observations be {(5,6), (5,5), (2,4), (2,6), (5,6), (2,9), (3,7), (10,4), (7,8)}. We use median as the parameter. So step 1, we calculate median of attribute 1 and attribute 2 which are 5 and 6 respectively. We add three derived features half1 half2 and cluster marker respectively. Now for each attribute in each observation, we mark half1 as 'H1' if attribute 1 value is less than equal to 5; else we mark half1 as 'H2'. Similarly, for half2 we mark it as 'H1' if value of attribute2 is less than equal to 6 else we mark as H2. We enclose the result in Table 1. The enumeration can look extremely simple; however we still wanted to furnish it, for easy implementation of our algorithm.

**Table 1: QuartileClustering example**

| Attribute 1 | Attribute 2 | Half1 | Half2 | Cluster Marker |
|---|---|---|---|---|
| 5 | 6 | $H_1$ | $H_1$ | $H_1 H_1$ |
| 5 | 5 | $H_1$ | $H_1$ | $H_1 H_1$ |
| 2 | 4 | $H_1$ | $H_1$ | $H_1 H_1$ |
| 2 | 6 | $H_1$ | $H_1$ | $H_1 H_1$ |
| 5 | 6 | $H_1$ | $H_1$ | $H_1 H_1$ |
| 2 | 9 | $H_1$ | $H_2$ | $H_1 H_2$ |
| 3 | 7 | $H_1$ | $H_2$ | $H_1 H_2$ |
| 10 | 4 | $H_2$ | $H_1$ | $H_2 H_1$ |
| 7 | 8 | $H_2$ | $H_2$ | $H_2 H_2$ |

.

Now we group the observations by Cluster Marker. Below are the 4 cluster that are generated.

Cluster 1 = {(5, 6), (5, 5), (2, 4), (2, 6), (5,6) }
Cluster 2 = {(2, 9), (3, 7)}
Cluster 3 = {(10, 4)}
Cluster 4 = {(7, 8)}

## 5 EXPERIMENT SETUP AND STUDY

This section consist of three empirical studies
1. We discuss a small dataset, run QuartileClustering algorithm on the same. First we show the added meaning to the clusters as compared to K means. We compare the results against other algorithms like K Means, PAM, EM and CLARA, in terms of the goodness of the clusters that are generated.
2. We run QuartileClustering against Iris dataset which contains pre-classified data. We check if the clusters generated are pure or not.
3. In our third experiment, we show how the added meaning to each cluster can help in a task like outlier detection on a set of SQL Queries.

### 5.1 Country Dataset

**Dataset Description:** Our dataset [27] contains a set of 70 records. Each record corresponds to a country. All the records have three attributes; country name, birth rate and death rate. The birth rate and death rate have been expressed in percentages. We have taken two attributes death rate and birth rate for analysis.

Table 2 records the comparison with the above algorithms and *QuartileClustering* for the country dataset

**Table 2: Comparison with other algorithms**

| Algorithm | SSE(Mean) | SSE (Median) | DB |
|---|---|---|---|
| QuartileCluetring | 2772.99 | 3037.54 | 0.320455 |
| K-means | 2450.21 | 2626.117 | 0.21415 |
| EM | 2708.91 | 3052.92 | 0.3324 |
| PAM | 1766.75 | 1864.2 | 0.556959 |
| CLARA | 1976.09 | 1946.35 | 0.717023 |

If we analyze the results as listed in Table 2, we see QuartileClustering to produce very comparable results with other algorithms. Our argument has always been that this algorithm attaches better semantic meaning with the clusters and with no prohibitive time complexity. The indices above prove that we have achieved this semantic meaning without faring poorly in terms of traditional measures for evaluation of clustering algorithm.

We see in Table 2 K-means the only algorithm which has outperformed QuartileClustering in all the indices. So to keep our discussion focused, we compare semantics/meaning of the clusters with only K-means. Table 3 lists centroid of the four clusters as generated by K - means. The tool used for the same is R.

**Table 3: K Mean's centroid for country dataset**

| Attributes | Clust1 | Clust2 | Clust3 | Clust4 |
|---|---|---|---|---|
| BR | 42.58 | 24.66 | 18.13 | 16.45 |
| DR | 12.70 | 7.86 | 12.41 | 8.09 |

Arguably this does not attach any semantics with the clusters as these clusters have been generated as a set of clusters which minimizes SSE. If we look at the clusters generated in Table 4, a very clear semantics or identification of the clusters are attached. There is also no sense of ordering in the cluster Ids.

**Table 4: *QuartileClusterig* generated clusters**

| Cluster Ids | Semantics |
|---|---|
| 1 | Birth rate in bottom Half, Death Rate also in bottom Half |
| 2 | Birth rate in bottom Half, Death Rate a in top Half |
| 3 | Birth rate in top Half, Death Rate a in bottom Half |
| 4 | Birth rate in top Half, Death Rate a in top Half |

Therefore, it is much easier to identify countries having population control problem, robust healthcare system, aging population etc. Rather than the absolute distance, the groupings are done in terms of their relative positioning in top and bottom half. As discussed earlier, for finer clusters we can go with 'Quartile' parameter.

We look at few examples now; we represent each ob-servation as Country (Birthrate, Deathrate) format in the discussion.

**Case 1: Low Birth rate and Low Death Rate**
The cluster which is closer to this in K-means is the fourth cluster. Countries like JAPAN (17.3, 7), KOREA (14.8, 5.7) , BOLIVIA ( 17.4, 5.8) are kept in same cluster by both the algorithms. Interestingly a country like CZECHOSLOVAKIA (16.9, 9.5) is also kept in the same cluster by K-means because of the similar nature of the value. QueryClustering succeeds to identify the relatively higher death rate as compared to all countries. (The median value is 9.15). Same is the case with SWEDEN (14.8, 10.1).

**Case 2: High Birth rate and Low Death Rate**
Cluster 2 of K-means is closest to this concept. Both countries VIETNAM (23.4, 5.1) and PORTUGAL (23.5, 10.8) are in the same cluster, reason being both of them are at almost same Euclidian distances 9.2 and 9.97 from the cluster centroid. QuartileClustering succeeds in identifying Vietnam's relatively lower death rate and clusters them in different clusters.

**Case 3: Low Birth rate and High Death Rate**
Cluster 3 of K-means is closest to this category. NORWAY (17.5, 10), BRITAIN (18.2, 12.2), BELGIUM (17.1, 12.7), FRANCE (18.2, 11.7) fall in the same cluster for both the algorithms. However K-means puts FINLAND (18.1, 9.2) in a different cluster because it is closer to Cluster 4 than that of cluster 3. (Euclidian distance 10 compared to 4).

**Case 4: High Birth rate and High Death Rate**
Cluster 1 of K-means is closest to this concept. We see countries like VENEZUELA (42.8, 6.7), PANAMA (40.1, 8), JORDAN (46.3, 6.4) getting added to this cluster. So basically K-means fells to recognize their relatively moderate deathrate and puts them with extreme values like IVORY COAST (56.1, 33.1) and GHANA (55.8, 25.6). QuartileClustering however succeeds to put VENEZUELA, PANAMA, JORDAN in one cluster and IVORY COAST and GHANA in a separate cluster. Another example from this cluster is MOROCCO (46.1, 18.7) and CONGO (37.3, 8) they have similar distance 48.36 and 49.97 respectively from the center as a result they are in the same cluster. QuartileClustering succeeds in putting them in different cluster so while CONGO goes in cluster 3; MOROCCO finds its place in cluster 4.

By the help of the above examples, we can see a clear meaning that QuartileClustering attaches to each cluster. We also see that it is more robust to extreme values because of its reliance on quartile rather than mean. If we explore the results of K-means we can realize the dominance of birthrate attribute over death rate because of its relatively higher values. We can normalize them and K-Mean can yield better result on the normalized attributes. However in case of QuartileClustering, this is independent of the same as its basis is on medians and quartiles rather than on absolute value.

## 5.2 IRIS dataset

**Dataset Description:**
This is one of the famous datasets with pre-classified one fifty records for three varieties of Iris flowers [28]. Each flower class has fifty observations each. Each observation constitutes of four attributes which are petal height, petal width, and sepal height and sepal width respectively.

One objective of any clustering algorithm is to identify natural clusters, which motivated us to pick up data with natural classification. An easy way to prove the same was to run the algorithm with 3 clusters and check if all the clusters are pure i.e. contains only populations belonging to a single class. We defined an alternate task where we clustered the dataset using QuartileClustering and analyzed if they are pure clusters or how many of them can form pure clusters. We used quartiles instead of halves in this case and listed down each cluster with corresponding Setosa, Virginica and Versicolor population. Interestingly, though this could have generated a total of two fifty six clusters, it only generates thirty three clusters. Among them only two clusters have mixed population, rest all of them are pure clusters, an accuracy of 93.93%. In terms of number of observations, 87.33% belonged to pure cluster. Table 5 illustrates the only two mixed clusters generated by the algorithm out of a total of 33 clusters.

**Table 5: Mixed Clusters**

| Cluster Label | Iris-setosa | Iris-versicolor | Iris-virginica |
|---|---|---|---|
| 4133 | 0 | 5 | 4 |
| 4233 | 0 | 7 | 3 |

Below are the two observations from a mixed cluster as shown in Table 6. While one of them is a versicolor, the other one is a virginica.

**Table 6: Observations from mixed Clusters**

| Id | Sepal Height | Sepal Width | Petal height | Petal Width | Class |
|---|---|---|---|---|---|
| 55 | 6.5 | 2.8 | 4.6 | 1.5 | Iris-versicolor |
| 124 | 6.3 | 2.7 | 4.9 | 1.8 | Iris-virginica |

If we analyze we will see the above two observations are really close to each other. We run the same dataset using K-Means and a total of thirty three clusters. Table 7, summarizes comparison with K-Mean and QuartileClustering.

**Table 7: Comparison with K Means**

| Algorithm | Pure Cluster (%) | Pure Population (%) |
|---|---|---|
| K Means | 87.87 | 87.33 |
| QuartileClustering | 93.93 | 87.33 |

The results show a bias towards QuartileClustering as we use the same no. of clusters for K Means. However in addition to maintaining the ability of detecting natural clusters very comparable with K Means, it brings the semantic meaning too. As an example when all the attributes are in top quartile it is a iris virginica.

## 5.3 SQL query dataset

**Dataset Description:**
This dataset consists of 26508 queries. Apart from the SQLQuery text, it has attributes like no. of rows returned, CPU cost, IO Cost, memory cost and run time. We have used this dataset extensively in [25] to identify outliers in SQL queries from performance tuning perspective. We have used various techniques like clustering, distance and density based methods to detect outliers to tune SQL queries. We run QuartileClustering on the same dataset and compare the results here. We used quartiles for the same. Our method generates a total of 255 clusters where the population in each cluster varies from 1 to 2192. When we compare this with the output of EM Algorithm, the results do not look converging. The notion we used for detecting outliers was to find outlier clusters i.e. clusters with lowest population and interestingly all these "outlying" queries turned out to be ill formed or costly queries.

Pragmatically enough, there are few queries in the dataset which are much outlying from the crowd. These are the ones that need immediate tuning for better system performance. If we critically evaluate how QuartileClustering faired here, at a high level, it does not seem very promising in terms of outlier detection. However it attaches the semantics with the clusters so we know a cluster like "Q4Q4Q4Q4" is the one we should probe from the perspective of query tuning, in fact, when we mix this with domain knowledge, we might also want to look at queries which fetches less no. of rows i.e not in top quartile as far as no. of rows is concerned but in other attributes it is in top quartile. We can immediately look at several such interesting combinations and achieve our end objective of identification of inefficient queries. We look at outliers generated by this algorithm for a couple of clusters and they appear to be absolutely normal and innocuous queries. Some using a select * with a like predicate and some doing simple selection from a table. So are they our point of interest? The answer to this is 'No', for this dataset, as the end objective was identification of inefficient queries by employing outlier detection method. These queries are not inefficient or costly because these are member of clusters like Q1Q2Q2Q3Q3. We argue our algorithm can attach semantic meaning to the clusters which helps to I) Identify clusters of inefficient queries II) It gives a generic idea on the distribution of the queries. In essence, we are better equipped to identify inefficient queries using this particular methodology. Another interesting byproduct of this algorithm is revelation of invalid combinations. This may not be an unknown knowledge but still this can systematically list series of invalid combinations. As example there is no member of Q1Q1Q1Q1Q4 cluster. So we can say there are very rare/no queries which return relatively less row, take relatively less time, consume relatively less CPU and IO but still need high memory for processing. This can also help to determine association among the attributes.

## 6 QUARTILECLUSTERING EVALUATION

When we summarize our findings of the experiments we can conclude as follows. I) this algorithm indeed attaches meaning to the clusters, which can be very useful to business users and domain experts rather than clusters defined by data boundaries, centroid a) without compromising on traditional clustering algorithms parameters like SSE or DB b) without compromising on the ability to identify natural clusters. II) We also evaluated this for outlier detection. The added semantics would be a worthy addition even in achieving this task. Also this can unearth interesting knowledge on distribution of the data, as example unavailability of few clusters can indicate some kind of association between the attributes.

Now coming to limitations, when we are doing clustering on a high dimensional dataset, it can generate very high no. of clusters. At a quartile level, for six parameters it has the potential to generate 4096 clusters. Even with the argument of added semantics, these many clusters will clutter business decisions and will be counter appealing. We would defend it saying the appeal of this algorithm will be more to the business users. For real word problems where meaning of the attribute plays a role, the no. of dimensions should not be very high except very specialized data in text mining or bio informatics. In fact [2] lists that 33% data mining tasks involve fewer than 10 variables. Another solution is that we can come up with some kind of merging without doing away with the semantics. Without digressing we would also like to add that traditional dimensional reduction techniques like Principle Component Analysis (PCA), Factor Analysis etc. would not have helped because we would have lost the semantics as components would have been some combination of the original attributes. Another limitation is that it does not cater to categorical attributes in its present form.

### 6.1 Merging Clusters

In this section, we define a rough scheme and direction for merging clusters. In fact, if we classify clustering algorithms, as partitioned and hierarchical, hierarchical algorithms naturally looks at merging clusters based on some metrics. However en route our journey to adding semantics, we do away with distance or similarity metrics. So we need to approach this in a different way, firstly how do we measure clusters which are closer? We take the difference in the quartile for selecting candidates for merging. So Q1Q1 is near to Q1Q2 than Q3Q1. We take absolute difference in the quartile no. for all attributes and then add it up. In case of a tie, we can calculate the mean and see which one is nearer. So a cluster Q1Q2Q3Q2 is at a distance (1+0+2+2) which is 5 with Q2Q2Q1Q4. This can also give rise to a tricky problem, after merging how we represent the merged cluster without losing on the semantics. So let's say we have two clusters as Q1Q2 and Q1Q1. We cannot add up the subscripts and represent the same as Q1Q3 because there is a cluster with the same literal strings even before the merge. So when we merge, we represent it as Q1H1, which makes perfect sense and we combine Q1, Q2 and Q3, Q4 rather than Q2, Q3, as bottom half and top half would be more meaningful then middle half. Similarly Q1H1 and Q1H2 can also be merged to represent as Q1F.

This raises another question, how we calculate distance between Q1H1 and Q1Q4? We cans take it as both 2 and 3. But the same thing we need to maintain across all merges.

A criticism of the same can be that we blur importance of individual attributes here. We can potentially use entropy or any other similar measure for a differential treatment. This is an area we would work further and run it through datasets. We can selectively use a mixed grain as well when adding differential treatment. So depending on the amount of information, diversity in an attribute we can choose to use halves for one attributes and quartile for another. We may also chose to go a level further like deciles for specific attributes. One evaluation technique for clustering algorithms is, if the algorithm detects clusters even if there is none present in the data. QuartileClustering in its present form is not equipped to handle this problem.

## 7 CONCLUSION

In this paper we have proposed a clustering / data partitioning methodology based on quartiles and medians. We demonstrate how this attaches meaning to each cluster. The same can be very helpful to business analysts to solve a business problem. We also establish through our experiments that it brings in this semantics without compromising on the ability to select clusters, or any other measure. The time-complexity of the algorithm in its natural form is linear. We argue that because of the natural meaning associated with the clusters and the simple implementation, it can be an ideal candidate for commercial implementations. We also see how this can aid a task like outlier detection by adding semantics to the clusters. Another by product of this algorithm is discovering the invalid combinations. This can definitely add some light to the inter-relationship of this attributes.

The algorithm suffers from the limitation that it gives rise to a high no. of clusters. We provide a rough direction on merging of clusters without losing on semantics. However this area needs to be researched further. In its present form, this only caters to numeric attributes. So as an extension, we need to work with categorical and binary attributes.

Overall we feel this algorithm can be really effective to make clusters more meaningful. It's linear time complexity and simple implementation adds to its advantage. The results are quite encouraging for the experiments conducted.

### REFERENCES

[1]  V Chandola, A Banerjee, B Kuman, Anomaly Detection: A Survey, ACM Computing Survey 2009.

[2]  Karl Rexer, "2010 survey summary report", 4th Annual Data Miner Survey, Rexer Analysis.

[3]  P. Berkhin. Survey of clustering data mining techniques. Technical report, Accrue Software, San Jose,CA, 2002.

[4]    Dunham, M. (2003). Data mining: introductory and advanced topics. Upper Saddle River, NJ: Prentice Hall.

[5]    Jain AK: Data clustering: 50 years beyond K-means. Pattern Recognition Letters 2010.

[6]    Tian Zhang, Raghu Ramakrishnan and Miron Livny, BIRCH: A New Data Clustering Algorithm and Its Applications

[7]    BIRCH: A New Data Clustering Algorithm and Its Applications , Data Mining and Knowledge Discovery, Mining and Knowledge Discovery , Volume 1, Number 2, 141-182, DOI: 10.1023/A:1009783824328

[8]    Karypis G., Han E. H. and Kumar V. (1999), CHAMELEON: A hierarchical clustering algorithm using dynamic modeling, Computer 32(8): 68-75, 1999.

[9]    Guha, S., Rastogi, R., Shim K. (1998), "CURE: An Efficient Clustering Algorithm for Large Data Sets", Published in the Proceedings of the ACM SIGMOD Conference.

[10]   Ester, M., Kriegel, H.-P., Sander, J., and Xu X. (1996), A density-based algorithm for discovering clusters in large spatial data sets with noise. Proc. 2nd Int. Conf. on Knowledge Discovery and Data Mining. Portland, OR, pp.226–231.

[11]   Guha, S, Rastogi, R., Shim K. (1999),"ROCK: A Robust Clustering Algorithm for Categorical Attributes", In the Proceedings of the IEEE Conference on Data Engineering.

[12]   Sheikholeslami, C., Chatterjee, S., Zhang, A.(1998), "WaveCluster: A-MultiResolution Clustering Approach for Very Large Spatial Data set". Proc. of 24th VLDB Conference.

[13]   J. Hartigan and M. Wong. Algorithm as136: A k-means clustering algorithm. Applied Statistics, 28:100–108, 1979.

[14]   Mianwei Zhou, Shenghua Bao, Xian Wu and Yong Yu, "An Unsupervised Model for Exploring Hierarchical Semantics from Social Annotations", Lecture Notes in Computer Science, 2007, Volume 4825/2007, 680-693.

[15]   Choudhury, B. and Bhattacharyya, P. (2002), "Text clustering using semantics," Proceedings of the 11thInternational World Wide Web Conference, WWW2002, Honolulu, Hawaii, USA, http://www2002.org/CDROM/poster/79.pdf.

[16]   John S. Zdanowicz, "Detecting Money Laundering and Terrorist Financing via Data Mining", COMMUNICATIONS OF THE ACM, May 2004/Vol. 47, No. 5.

[17]   Randall E. Duran, Li Zhang and Tom Hayhurst , "Applying Soft Cluster Analysis Techniques to Customer Interaction Information" , Studies in Fuzziness and Soft Computing, 2010, Volume 258/2010, 49-78.

[18]   Mehmed Kantardzic, "Data Mining—Concepts, Models, Methods, and Algorithms "

[19]   A. K. Jain, M. N. Murty, and P. J. Flynn. Data clustering: A review. ACM Comput. Surv., 31(3):264–323, 1999.

[20]   R. Xu and D. Wunsch, "Survey of clustering algorithms," IEEE Trans. Neural Networks, no. Vol.16, Issue 3, pp. 645– 678, May 2005.

[21]   Hawkins D, "Identification of Outliers", Chapman and Hall, 1980.

[22]   J. Hodge (vicky@cs.york.ac.uk)   and Jim Austin , "A Survey of Outlier Detection Methodologies" Victoria Artificial Intelligence Review, 2004.

[23]   Kaufman, Leonard, Rousseeuw, Peter J, "Finding groups in data: An introduction to cluster analysis", Wiley series in probability and statistics, 2005.

[24]   Samiran Ghosh, Saptarsi Goswami, Amlan Chakrabarti , "Outlier Detection Techniques for SQL and ETL Tuning", International Journal of Computer Applications , Volume 23– No.8, June 2011.

[25]   http://archive.ics.uci.edu/ml/datasets/Iris, Contributed by Fisher, 1936

[26]   http://people.sc.fsu.edu/~jburkardt/datasets/hartigan/file26.txt, John Hartigan

[27]   Coronary calcium and cardiovascular event risk: evaluation by age-and sex-specific quartile, Nathan D. Wong, PhD, Matthew J. Budoff, MD, Jose Pio, and Robert C. Detrano, MD, PhD Irvine and Torrance, Calif, American heart journal, 2002

[28]   Vitamin Intake and Risk of Alzheimer Disease, Jose A. Luchsinger, MD; Ming-Xin Tang, PhD; Steven Shea, MD; Richard Mayeux, MD , Arch Neurol. 2003;60:203-208

**S. Goswami** received the B.Tech in Electronics Engineering from MNIT, Jaipur in 2001. He has completed M.Tech (CS) from AKCSIT, Calcutta University. He is enrolled as a PHD student in the same institute .He has 9 + years of experience in software industry. He is currently leading a team of 80+ individuals in a Fortune 500 Company. His area of expertise includes data warehousing, business intelligence and data mining.

**Dr. A. Chakrabarti** is presently a Visiting Research Associate in the Dept. of Electrical Engineering, Princeton University, U.S.A., he also holds a permanent position of a Reader in the A.K.Choudhury School of Information Technology, University of Calcutta, India. He obtained his M.Tech. and Ph.D (Tech.) degrees from the University of Calcutta. During his Ph.D. he has done his research in the domain of Quantum Information Theory in collaboration with Indian Statistical Institute, Kolkata, India. He has been the recipient of BOYSCAST award from the Department of Science and Technology, Govt. of India, in the year 2011, for his research contribution in engineering. His present research interests are Quantum Computing, Machine Learning and Intelligent Systems, VLSI design, Embedded System Design and Video & Image Processing. He has around 40 research publications in International Journal and Conferences.